\documentclass[reprint,aps,pra,groupedaddress,superscriptaddress,amsmath,amssymb,
showpacs,showkeys]{revtex4-1}

\usepackage{dcolumn}
\usepackage{bm}
\usepackage[mathlines]{lineno}
\usepackage{amssymb,epsfig}
\usepackage{color}

\begin{document}
\title{Generalized James' effective Hamiltonian method}
\author{Wenjun Shao}
\affiliation{Department of Physics, Shanghai Normal University, Shanghai
200234, China}
\author{Chunfeng Wu}
\affiliation{Pillar of Engineering Product Development, Singapore University of Technology
and Design, 8 Somapah Road, Singapore 487372 }
\author{Xun-Li Feng}
\email{xlfeng@shnu.edu.cn}
\affiliation{Department of Physics, Shanghai Normal University, Shanghai
200234, China}

\begin{abstract}
James' effective Hamiltonian method has been extensively adopted to
investigate largely detuned interacting quantum systems. This method is just corresponding
to the second-order perturbation theory, and cannot be exploited to treat the
 problems which should be solved by using the third- or higher-order
perturbation theory. In this paper, we generalize James'
effective Hamiltonian method to the higher-order case. 
Using the method developed here, we reexamine
two examples published recently [Phys. Rev.Lett. \textbf{117}, 043601 (2016), 
Phys. Rev A \textbf{92}, 023842 (2015)], our results turn out to be
the same as the original ones derived from the third-order perturbation
theory and adiabatic elimination method respectively. For some specific
problems, this method can simplify the calculating procedure, and the
resultant effective Hamiltonian is more general.
\end{abstract}

\maketitle
\section{Introduction}

Generally it is hard to obtain exact analytical solutions for quantum systems,
thus various approximation methods have been developed to deal with couplings
between the quantum systems. For instance, perturbation theory \cite{QM},
adiabatic elimination method \cite{Eberly,Law} and James' effective
Hamiltonian method\ \cite{F-james,C-james,PRA-james} have been utilized in the
derivation of effective Hamiltonian in quantum mechanics. However, for some
specific physical problems the former two methods are time consuming and the
deriving procedure is tediously long. While James' effective Hamiltonian
method may sometimes provide us an efficient tool and has been employed to
solve many interesting questions related to light-matter interactions
\cite{MS-cavity,Plenio,F1}.

We have noted that the James' effective Hamiltonian method is actually
corresponding to the second-order perturbation theory. For example, the
effective Hamiltonian of the M\o lmer and S\o rensen\textbf{ }scheme for
trapped ions can be derived from both the second-order perturbation theory
\cite{MS1,MS2} and the James' effective Hamiltonian method \cite{F-james}. Accordingly in
what follows we refer to the original James' effective Hamiltonian method as
the second-order James' method.

Very recently, the third-order perturbation theory was exploited to deal with
problems related to the ultrastrong coupling between the cavity field and
atoms. For instance, by applying the third-order perturbation theory, a
surprising result that one photon can simultaneously excite two or more atoms
was deduced in the ultrastrong coupling regime with a symmetry-broken
potential \cite{Nori}. As the other example, when the frequency of the cavity field is near
one-third of the atomic transition frequency, a resonant three-photon coupling
between a two-level atom and the cavity field can be constructed which was
derived by the adiabatic elimination method in Law's group \cite{Law}. We find
one can obtain the similar result by using the third-order perturbation
theory. Unfortunately, one cannot derive the effective Hamiltonian appearing
in Refs. \cite{Nori} and \cite{Law} by using original James' effective
Hamiltonian method, the reason is that James' method is only corresponding to
the second-order perturbation theory as mentioned above. So it is necessary to
generalize James' method to higher-order. This is the purpose of our present work.

The paper is organized as follows. In Sec. II, we give a simple review about
James' effective Hamiltonian method. And in Sec. III, we derive the
third- and $n$th-order James' effective Hamiltonian by generalizing James' method. Then
in Sec. IV we use the generalized James' method to derive the effective
Hamiltonian in Refs. \cite{Nori} and \cite{Law}, and compare both results.
Finally, a summary is made in Sec. V.

\section{Review of James' effective Hamiltonian method}

Now let us give a simple review about James' effective Hamiltonian method
\cite{C-james,PRA-james}, which is well suitable for treating strongly detuned
interacting quantum systems, such as atomic or molecular interacting with
laser beams, spin system interacting with oscillating magnetic fields under
the conditions of highly detuned regime. In the interaction picture the
system mentioned above is governed by the Hamiltonian, $\hat{H}_{I}\left(
t\right)  $, of the form%
\begin{equation}
\hat{H}_{I}\left(  t\right)  =\sum_{m}\left[  \hat{h}_{m}\exp\left(
i\omega_{m}t\right)  +\hat{h}_{m}^{^{\dagger}}\exp\left(  -i\omega
_{m}t\right)  \right]  . \label{1}%
\end{equation}
The Schr\"{o}dinger equation is%
\begin{equation}
i\hbar\frac{\partial}{\partial t}\left\vert \psi\left(  t\right)
\right\rangle =\hat{H}_{I}\left(  t\right)  \left\vert \psi\left(  t\right)
\right\rangle . \label{2}%
\end{equation}
The formal solution to Eq. ($2$) is%
\begin{equation}
\left\vert \psi\left(  t\right)  \right\rangle =\left\vert \psi\left(
0\right)  \right\rangle +\frac{1}{i\hbar}\int_{0}^{t}\hat{H}_{I}\left(
t^{\prime}\right)  \left\vert \psi\left(  t^{\prime}\right)  \right\rangle
dt^{\prime}. \label{3}%
\end{equation}
Substituting Eq. ($3$) into Eq. ($2$) and neglecting highly oscillating terms
$\hat{H}_{I}\left(  t\right)  \left\vert \psi\left(  0\right)  \right\rangle $
yield%
\begin{equation}
i\hbar\frac{\partial}{\partial t}\left\vert \psi\left(  t\right)
\right\rangle =\frac{1}{i\hbar}\hat{H}_{I}\left(  t\right)  \int_{0}^{t}%
\hat{H}_{I}\left(  t^{\prime}\right)  \left\vert \psi\left(  t^{\prime
}\right)  \right\rangle dt^{\prime}. \label{4}%
\end{equation}
By Markovian approximation, Eq. ($4$) becomes
\begin{equation}
i\hbar\frac{\partial}{\partial t}\left\vert \psi\left(  t\right)
\right\rangle =\hat{H}_{eff}^{\left(  2\right)  }\left(  t\right)  \left\vert
\psi\left(  t\right)  \right\rangle , \label{5}%
\end{equation}
where%
\begin{equation}
\hat{H}_{eff}^{\left(  2\right)  }\left(  t\right)  =\frac{1}{i\hbar}\hat
{H}_{I}\left(  t\right)  \int_{0}^{t}\hat{H}_{I}\left(  t^{\prime}\right)
dt^{\prime}. \label{6}%
\end{equation}
Here the superscript "$(2)$" indicates the second-order James' effective
Hamiltonian as mentioned in Sec. I, which is corresponding to the second-order
perturbation theory.

Substituting Eq. ($1$) into Eq. ($6$) and taking all frequencies $\omega_{m}$
distinct, we obtain the following form by using the rotating wave approximation

\begin{equation}
\hat{H}_{eff}^{\left(  2\right)  }\left(  t\right)  =\sum_{m}\frac{1}%
{\hbar\omega_{m}}\left[  \hat{h}_{m},\hat{h}_{m}^{^{\dagger}}\right].
\end{equation}

Note that for the case that the frequencies $\omega_{m}$ are not all distinct,
e.g., $\left\vert \omega_{m}-\omega_{n}\right\vert \ll\omega_{m},\omega_{n},$
one should take into account the terms containing $\hat{h}_{m}\hat{h}%
_{n}^{^{\dagger}}e^{i\left(  \omega_{m}-\omega_{n}\right)  t}$, $\hat{h}%
_{m}^{^{\dagger}}\hat{h}_{n}e^{-i\left(  \omega_{m}-\omega_{n}\right)  t}$ as
did in \cite{C-james}. However, for simplicity here and in what follows we
only consider the case of all the frequencies $\omega_{m}$ being distinct.

\section{Generalized James' effective Hamiltonian method}

In order to develop a generalized James' method corresponding to the third-
and higher-order perturbation theory, we adopt James' idea as mentioned in
Sec. II. First of all, substituting Eq. ($3$) into Eq. ($4$) with iteration and yields%
\begin{widetext}
\begin{align}
i\hbar\frac{\partial}{\partial t}\left\vert \psi\left(  t\right)
\right\rangle &= \left[ \frac{1}{i\hbar}\hat{H}_{I}\left(  t\right)  \int_{0}^{t}%
\hat{H}_{I}\left(  t_{1}\right) dt_{1}+ \left( \frac{1}{i\hbar}\right)^{2}
 \hat{H}_{I}\left(  t\right)  \int_{0}^{t}%
\hat{H}_{I}\left(  t_{1}\right)  \int_{0}^{t_{1}}\hat{H}_{I}\left(
t_{2} \right) dt_{2}  dt_{1} + \cdots  \right. \nonumber\\
&  \left.   +\left( \frac{1}{i\hbar}\right)^{n-1}
\hat{H}_{I}\left(  t\right)  \int_{0}^{t}\hat{H}_{I}\left(  t_{1}\right) 
\int_{0}^{t_{1}}\hat{H}_{I}\left(  t_{2}\right) \cdots    
\int_{0}^{t_{n-2}}\hat{H}_{I}\left(  t_{n-1}\right)dt_{n-1} 
\cdot\cdot\cdot dt_{2} dt_{1} + \cdots \right] \left\vert \psi\left(  0\right)
\right\rangle,  \label{8}%
\end{align}
\end{widetext}
Taking Markovian approximation, we obtain%
\begin{equation}
i\hbar\frac{\partial}{\partial t}\left\vert \psi\left(  t\right)
\right\rangle =\hat{H}_{eff} \left(  t\right)\left\vert
\psi\left(  t\right)  \right\rangle,
\end{equation}%
where
\begin{equation}
\hat{H}_{eff}\left(  t\right)=\hat{H}_{eff}^{\left(  2\right)  }\left(  t\right)
+\hat{H}_{eff}^{\left(3\right)  }\left(  t\right)+\cdots
+\hat{H}_{eff}^{\left(  n\right)  }\left(  t\right)+\cdots ,\label{10}%
\end{equation}%
$\hat{H}_{eff}^{\left(  2\right)  }\left(  t\right)  $ is the
second-order James' effective Hamiltonian expressed by Eq. ($6$) and $\hat
{H}_{eff}^{\left(  3\right)  }\left(  t\right)  $ is referred to as the
third-order James' effective\ Hamiltonian of the following form%
\begin{equation}
\hat{H}_{eff}^{\left(  3\right)  }\left(  t\right)  =-\frac{1}{\hbar^{2}}%
\hat{H}_{I}\left(  t\right)  \int_{0}^{t}\hat{H}_{I}\left(  t_{1}\right)
\int_{0}^{t_{1}}\hat{H}_{I}\left(  t_{2}\right) dt_{2} dt_{1},
\end{equation}
$\hat{H}_{eff}^{\left(  n\right)  }\left(  t\right) $ is the $n$th-order James' effective Hamiltonian 
\begin{align}
\hat{H}_{eff}^{\left(  n\right)  }\left(  t\right)  =& \left( \frac{1}{i\hbar}\right)^{n-1}
\hat{H}_{I}\left(  t\right)  \int_{0}^{t}\hat{H}_{I}\left(  t_{1}\right) 
\int_{0}^{t_{1}}\hat{H}_{I}\left(  t_{2}\right) \times  \nonumber\\
&\cdots \int_{0}^{t_{n-2}}\hat{H}_{I}\left(  t_{n-1}\right)dt_{n-1} 
\cdots dt_{2} dt_{1}. \label{12}
\end{align}

Examining our results (10)--(12), one can find the effective Hamiltonian in arbitrary 
orders is actually equivalent to the series expansion of the unitary evolution 
operator $U(t,0)$ \cite{De,Louisell}. To this end, now let us briefly derive the 
series expansion of the unitary evolution operator from  Eq. (8). The formal solution 
to Eq. (8) is
\begin{align}
\left\vert \psi\left(  t\right)  \right\rangle &=\left[ 1+\frac{1}{i\hbar}\int_{0}^{\tau}
\hat{H}_{eff} \left(  t\right)dt \right] \left\vert\psi\left( 0\right) 
 \right\rangle \nonumber\\
& = U(t,0) \left\vert\psi\left( 0\right)  \right\rangle ,
\end{align}
where
\begin{align}
U(t,0)&=1+\sum_{n=1}^{\infty }\left( \frac{1}{i\hbar}\right)^{n}  \int_{0}^{\tau}
\hat{H}_{I}\left(  t\right)  \int_{0}^{t}\hat{H}_{I}\left(  t_{1}\right) \times \nonumber\\
&\cdots \int_{0}^{t_{n-2}}\hat{H}_{I}\left(  t_{n-1}\right)dt_{n-1} 
\cdots dt_{1}dt.
\end{align}
It is not difficult to check that $U(t,0)$ in Eq. ($14$) is exactly the series expansion 
of the unitary evolution operator \cite{De,Louisell}. Different from the method of 
series expansion of the unitary evolution operator, the generalized James' method 
developed here can give directly the effective Hamiltonian in arbitrary order.

In what follows we mainly focus on the third-order case and 
limit ourselves to the case that all of the frequencies $\omega_{m}$ are not
only distinct, but also the algebraic sum of any three frequencies including 
two same ones is zero or distinct from zero.
 Considering $\hat{H}_{I}\left(  t\right)  $ taking the
form of Eq. ($1$), one can further simplify 
$\hat{H}_{eff}^{\left(  3\right)  }\left(  t\right)  $ by using the rotating
wave approximation,%
\begin{widetext}
\begin{align}
\hat{H}_{eff}^{\left(  3\right)  }\left(  t\right)   &  =\frac{1}{\hbar^{2}%
}\sum_{l,m,n}\left\{  \frac{1}{\omega_{n}\left(  \omega_{n}-\omega_{m}\right)
}\left[  \hat{h}_{l}\hat{h}_{m}^{^{\dagger}}\hat{h}_{n}e^{i\left(  \omega
_{l}-\omega_{m}+\omega_{n}\right)  t}+\hat{h}_{l}^{^{\dagger}}\hat{h}_{m}%
\hat{h}_{n}^{^{\dagger}}e^{i\left(  -\omega_{l}+\omega_{m}-\omega_{n}\right)
t}+\hat{h}_{l}\hat{h}_{m}\hat{h}_{n}^{^{\dagger}}e^{i\left(  \omega_{l}%
+\omega_{m}-\omega_{n}\right)  t}\right.  \right. \nonumber\\
&  \left. +\hat{h}_{l}^{^{\dagger}}\hat{h}%
_{m}^{^{\dagger}}\hat{h}_{n}e^{i\left(  -\omega_{l}-\omega_{m}+\omega
_{n}\right)  t}\right] \left.  +\frac{1}{\omega_{n}\left(  \omega_{n}+\omega_{m}\right)  }\left[
\hat{h}_{l}^{^{\dagger}}\hat{h}_{m}\hat{h}_{n}e^{i\left(  -\omega_{l}%
+\omega_{m}+\omega_{n}\right)  t}+\hat{h}_{l}\hat{h}_{m}^{^{\dagger}}\hat
{h}_{n}^{^{\dagger}}e^{i\left(  \omega_{l}-\omega_{m}-\omega_{n}\right)
t}\right]  \right\}  . \label{15}%
\end{align}
\end{widetext}

Note that in the above equation only the terms with sum frequency $\omega
_{l}+\omega_{m}+\omega_{n}$ have been neglected. Since the frequencies
$\omega_{m}$ are all distinct, one needs only to keep the terms in $\hat
{H}_{eff}^{\left(  3\right)  }$ from the contributions of algebraic sum of any
three frequencies being zero and other contributions are all neglected
according to rotating wave approximation. That is to say, Eq. ($15$) can be
further simplified according to the practical situation. After such
simplification, it is not difficult to prove $\hat{H}_{eff}^{\left(  3\right)
}$ is hermitian, which is provided in appendix.

\section{Examples}

In this section, we reexamine two examples in Refs. \cite{Nori} and \cite{Law}
by the generalized James' method. As the first example, we examine two (or
more) atoms excited simultaneously by one photon and compare 
the results with the original ones from the third
perturbation theory in \cite{Nori}. Then we revisit three-photon absorption of
a two-level atom via the counter-rotating processes in the Rabi model
\cite{Law}.

\subsection{\textbf{Two atoms excited simultaneously by one photon }}

In a recent literature an interesting result that two or more atoms can be
simultaneously excited by one photon was derived in the ultrastrong coupling
regime with a symmetry-broken potential \cite{Nori}. The system considered in
\cite{Nori} is two or more identical qubits coupling to a single cavity mode.
In the following we apply the generalized James' method to revisit this
question in the case of two atoms. In the interaction picture with respect to
$\hat{H}_{0}=\frac{1}{2}\omega_{q}\sum_{i=1,2}\hat{\sigma}_{z}^{i}+\omega
_{c}a^{\dagger}a$ the Hamiltonian is given by $\left(  \hbar=1\right)  $%
\begin{align}
\hat{H}_{I}  & =\sum_{i=1,2}\lambda\hat{a}^{\dagger}\left[  \cos\theta\left(
\hat{\sigma}_{-}^{i}e^{i\omega_{q}t}+\hat{\sigma}_{+}^{i}e^{3i\omega_{q}%
t}\right)  \right.  \nonumber\\
& \left.  +\sin\theta\hat{\sigma}_{z}^{i}e^{2i\omega_{q}t}\right]
+H.c,\label{16}%
\end{align}
where $\omega_{c}$ ($\omega_{q}$) is the resonant frequency of the cavity mode
(the qubit transition frequency), $\hat{\sigma}_{x}^{i}$ and $\hat{\sigma}%
_{z}^{i}$ are Pauli operators for the $i$th qubit, $a^{\dagger}$ ($a$) is the
photon creation (annihilation) operator for cavity mode, and $\lambda$ is the
coupling strength of each qubit to the cavity mode. Note that to get Eq. ($16$) 
we have taken $\omega_{c}=2\omega_{q}$. According to Eq. ($1$), 
$\hat{H}_{I}$ expressed in Eq. ($16$) possesses three distinct frequencies,
$\omega_{1}=\omega_{q}$, $\omega_{2}=2\omega_{q}$ and $\omega_{3}=3\omega_{q}$
and the corresponding $\hat{h}_{m}$ $\left(  m=1,2,3\right)  $ are of the
form
\begin{align*}
\hat{h}_{1}  &  =\lambda\sum_{i=1,2}\cos\theta\hat{a}^{\dagger}\hat{\sigma
}_{-}^{i},\\
\hat{h}_{2}  &  =\lambda\sum_{i=1,2}\sin\theta\hat{a}^{\dagger}\hat{\sigma
}_{Z}^{i},\\
\hat{h}_{3}  &  =\lambda\sum_{i=1,2}\cos\theta\hat{a}^{\dagger}\hat{\sigma
}_{+}^{i}.
\end{align*}
By utilizing the third-order James' effective Hamiltonian method, we arrive at
\begin{equation}
\hat{H}_{eff}^{\left(  3\right)  }=-\frac{8\lambda^{3}\cos^{2}\theta\sin
\theta}{3\omega_{q}^{2}}\left(  \hat{a}^{\dagger}\hat{\sigma}_{-}^{1}%
\hat{\sigma}_{-}^{2}+\hat{a}\hat{\sigma}_{+}^{1}\hat{\sigma}_{+}^{2}\right)  .
\label{14}%
\end{equation}

First of all, let us assume the initial state of the system is $\left\vert
gg1\right\rangle $ as did in Ref. \cite{Nori}, where $\left\vert
gg1\right\rangle \equiv\left\vert g\right\rangle _{1}\otimes\left\vert
g\right\rangle _{2}\otimes\left\vert 1\right\rangle _{C}$ standing for both
atoms are in their ground state and the cavity mode in the one photon Fock
state. In this case $\hat{H}_{eff}^{\left(  3\right)  }$ turns out to be the
same as that in Ref. \cite{Nori}. Obviously, our method is much simpler than
the third perturbation theory used in \cite{Nori}.

Moreover, Eq. ($17$) shows richer physical connotation if one considers the
initial state of the cavity mode in Fock state $\left\vert n\right\rangle
_{C}$ and both atoms are still in their ground state, in such a case, Eq.
($17$) can be written as
\begin{equation}
\hat{H}_{eff}^{\left(  3\right)  }=\Omega_{eff}^{\left(  3\right)  }\left(
\left\vert ee\left(  n-1\right)  \right\rangle \left\langle ggn\right\vert
+\left\vert ggn\right\rangle \left\langle ee\left(  n-1\right)  \right\vert
\right)  ,
\end{equation}
where $\Omega_{eff}^{\left(  3\right)  }=-\frac{8\sqrt{n}\lambda^{3}\cos
^{2}\theta\sin\theta}{3\omega_{q}^{2}}$ is the effective Rabi frequency, and
it is proportional to $\sqrt{n}$. That is to say, more photons in the cavity
can enhance the ability for one photon to simultaneously excite two atoms.

\subsection{\textbf{Three-photon coupling in the large-detuned Rabi model}}

In the recent contribution, Ma and Law investigated theoretically the
three-photon coupling in Rabi model in the large-detuning regime, they found a
two-level atom can absorb three photons simultaneously via the
counter-rotating processes in the three-photon resonance \cite{Law}. In their
study, the adiabatic elimination method was used to derive the effective
Hamiltonian. Here we show the generalized James' method is also available to
get the same result.

In the interaction picture with respect to $\hat{H}_{0}=\frac{1}{2}\omega
_{a}\hat{\sigma}_{z}+\omega_{c}\hat{a}^{\dagger}\hat{a}$ the Hamiltonian of
the quantum Rabi model $\left(  \hbar=1\right)  $ is given by%
\begin{equation}
\hat{H}_{I}=\lambda\left[  \hat{a}e^{i\left(  \omega
_{a}-\omega_{c}\right)  t}+\hat{a}^{\dagger}e^{i\left(  \omega_{a}+\omega
_{c}\right)  t}\right]\hat{\sigma}_{+}  +H.c,
\end{equation}
where $\lambda$ is the coupling rate of atom to cavity mode, $\hat{a}$ and
$\hat{a}^{\dagger}$ are, respectively, the annihilation and creation operators
for the cavity field of frequency $\omega_{c}$, and $\omega_{a}$ is the atom
transition frequency. The Pauli matrices are defined as $\hat{\sigma}%
_{z}=\left\vert e\right\rangle \left\langle e\right\vert -\left\vert
g\right\rangle \left\langle g\right\vert ,$ $\hat{\sigma}_{+}=\left\vert
e\right\rangle \left\langle g\right\vert $, $\hat{\sigma}_{-}=\left\vert
g\right\rangle \left\langle e\right\vert $ and$\ \hat{\sigma}_{x}=\hat{\sigma
}_{+}+\hat{\sigma}_{-}$. Under the three-photon resonance with $\omega
_{c}=\omega_{a}/3$, the interaction Hamiltonian becomes%
\begin{equation}
\hat{H}_{I}=\lambda\hat{\sigma}_{+}\left(  \hat{a}e^{2i\omega_{c}t}+\hat
{a}^{\dagger}e^{4i\omega_{c}t}\right)  +H.c.
\end{equation}

Making the identification $\hat{h}_{1}=\lambda\hat{a}\hat{\sigma}_{+}$ with
frequency $\omega_{1}=2\omega_{c},$ $\hat{h}_{2}=\lambda\hat{a}^{\dagger}%
\hat{\sigma}_{+}$ with frequency $\omega_{2}=4\omega_{c}$, one can
straightforwardly utilize the formula expressed in Eq. ($10$) to find
the effective Hamiltonian%
\begin{align}
\hat{H}_{eff}^{\left(  2\right)  }  &  =\frac{\lambda^{2}}{4\omega_{c}}\left[
\left(  3\hat{a}^{\dagger}\hat{a}+2\right)  \hat{\sigma}_{+}\hat{\sigma}%
_{-}-\left(  3\hat{a}^{\dagger}\hat{a}+1\right)  \hat{\sigma}_{-}\hat{\sigma
}_{+}\right]  ,\\
\hat{H}_{eff}^{\left(  3\right)  }  &  =-\frac{\lambda^{3}}{4\omega_{c}^{2}%
}\left[  \left(  \hat{a}^{\dagger}\right)  ^{3}\hat{\sigma}_{-}+\hat{a}%
^{3}\hat{\sigma}_{+}\right]  .
\end{align}

If we take the specific state $\left\vert g,3\right\rangle $ as the initial
state of the system as did in Ref. \cite{Law} and consider in the same
picture, say, interaction picture, the Eqs. ($21$) and ($22$) turn out to be the same as
the result obtained in \cite{Law}. Moreover, if we take $\left\vert
g,n\right\rangle \ $as the initial state, the Eqs. ($21$) and ($22$) can give a 
more general result.

\section{Summary}

In this work, we have generalized James' effective Hamiltonian method which
corresponds to the second-order perturbation theory to the case corresponding
to the third- and higher-order perturbation theory, and we have shown that the 
effective Hamiltonian in arbitrary orders developed is actually equivalent to 
the series expansion of the unitary evolution operator. By using the generalized James'
effective Hamiltonian method, we have reexamined two examples \cite{Nori,Law}
published recently, the resultant Hamiltonians are the same as the original ones derived from
the third-order perturbation theory and adiabatic elimination method
respectively. The generalized James' effective Hamiltonian method developed
here can not only simplify the calculating procedure for some problems, but
also provide us richer and more general results. We hope the generalized
James' effective Hamiltonian method can be applicable to solve more quantum questions.

\section*{ACKNOWLEDGMENT}

We are grateful to Yu Sixia for the helpful discussion and to the anonymous reviewer 
for the constructive suggestions and providing the Ref. \cite{De}. This work is supported
by the Natural Science Foundation of Shanghai (Grant No. 15ZR1430600), National 
Natural Science Foundation of China under Grant Nos. 61475168, 11674231 and
11074079. XLF is sponsored by Shanghai Gaofeng \& Gaoyuan Project for
University Academic Program Development.

\section*{APPENDIX: proving hermiticity of $\hat{H}_{eff}^{\left(  3\right)  }%
$}

In this appendix we will prove the hermiticity of $\hat{H}_{eff}^{\left(
3\right)  }$. For simplicity, we concentrate our attention on the case, as
mentioned in the text, that all of the frequencies $\omega_{m}$ are distinct,
and the algebraic sum of any three frequencies is zero or distinct from zero, 
which includes both of the three frequencies are the same. Under such
conditions, $\hat{H}_{eff}^{\left(  3\right)  }$ only contains the terms from
the contributions of algebraic sum of any three frequencies being zero and
other contributions are all neglected according to rotating wave
approximation. Apparently it is sufficient to prove one of such terms is
hermitian. Without loss of generality, we suppose 
$\omega_{l}$, $\omega_{m}$ and $\omega_{n}$ are such
three frequencies satisfying $\omega_{l}+\omega_{m}-\omega_{n}=0$, their
contribution in $\hat{H}_{eff}^{\left(  3\right)  }$ is set to $V_{lmn}$,
$V_{lmn}$ can be simplified by using $\omega_{l}=\omega_{n}-\omega_{m}$
\begin{widetext}
\begin{align}
V_{lmn}=\frac{1}{\hbar^{2}}\left[  \frac{\hat{h}_{n}^{^{\dagger}}\hat{h}%
_{l}\hat{h}_{m}+\hat{h}_{n}\hat{h}_{l}^{^{\dagger}}\hat{h}_{m}^{\dagger}%
}{\omega_{n}\omega_{m}}+\frac{\hat{h}_{m}\hat{h}_{n}^{^{\dagger}}\hat{h}%
_{l}+\hat{h}_{m}^{\dagger}\hat{h}_{n}\hat{h}_{l}^{^{\dagger}}}{\omega
_{m}\left(  \omega_{m}-\omega_{n}\right)  }+\frac{\hat{h}_{l}\hat{h}_{m}%
\hat{h}_{n}^{^{\dagger}}+\hat{h}_{l}^{^{\dagger}}\hat{h}_{m}^{\dagger}\hat
{h}_{n}}{\omega_{n}\left(  \omega_{n}-\omega_{m}\right)  }\right]  +H.c.
\end{align}
\end{widetext}

It is not difficult to find that Eq. ($20$) is hermitian. For the case of
$-\omega_{l}-\omega_{m}+\omega_{n}=0$, we can also prove its hermiticity. If
two of the three frequencies are the same, say, $\omega_{l}=\omega_{m}$, one
can prove this is just a special case.\ Therefore, we can ensure that the
third-order James' effective Hamiltonian $\hat{H}_{eff}^{\left(  3\right)  }$
is hermitian.


\begin{thebibliography}{99}                                                                                               

\bibitem {QM}L. D. Landau, E. M. Lifshitz, \textit{Quantum Mechanics: Non-Relativistic
Theory} (Butterworth-Heinemann, 1981), Chap. 6.

\bibitem {Eberly}L. Wang, R. R. Puri, and J. H. Eberly, Coupled-channel cavity
QED model and exact solutions, Phys. Rev. A \textbf{46}, 7192 (1992).

\bibitem {Law}Ken K. W. Ma and C. K. Law, Three-photon resonance and adiabatic
passage in the large-detuning Rabi model, Phys. Rev A \textbf{92}, 023842 (2015).

\bibitem {F-james}D. F. V. James, Quantum Computation with Hot and Cold Ions:
An Assessment of Proposed Schemes, Fortschr. Phys. \textbf{48}, 823 (2000).

\bibitem {C-james}D. F. V. James, J. Jerke, Effective Hamiltonian Theory and
Its Applications in Quantum Information, Can. J. Phys. \textbf{85}, 625 (2007).

\bibitem {PRA-james}O. Gamel and D. F. V. James, Time-averaged quantum
dynamics and the validity of the effective Hamiltonian model, Phys. Rev. A
\textbf{82}, 052106 (2010).

\bibitem {MS-cavity}A. S\o rensen and K. M\o lmer, Entangling atoms in bad
cavities, Phys. Rev. A \textbf{66}, 022314 (2002).

\bibitem {Plenio}M. J. Hartmann, F. G. S. L. Brandao, and M. B. Plenio,
Effective Spin Systems in Coupled Microcavities, Phys. Rev. Lett. \textbf{99},
160501 (2007).

\bibitem {F1}X. L Feng, C. Wu, H. Sun, and C. H. Oh, Geometric Entangling
Gates in Decoherence-Free Subspaces with Minimal Requirements, Phys. Rev.
Lett. \textbf{103}, 200501 (2009).

\bibitem {MS1}A. S\o rensen and K. M\o lmer, Quantum Computation with Ions in
Thermal Motion, Phys. Rev. Lett. \textbf{82}, 1971 (1999).

\bibitem {MS2}K. M\o lmer and A. S\o rensen, Multiparticle Entanglement of Hot
Trapped Ions, Phys. Rev. Lett. \textbf{82}, 1835 (1999).

\bibitem {Nori}L. Garziano, V. Macr, R. Stassi, O. Di Stefano, F. Nori, and S.
Savasta, One Photon Can Simultaneously Excite Two or More Atoms, Phys. Rev.
Lett. \textbf{117}, 043601 (2016).

\bibitem {De} A. L. Fetter and J. D. Waleka, \textit{Quantum theory of many-particle 
systems} (New York, 1971), Chap. 3, Eq. (6.19).

\bibitem {Louisell}W. H. Louisell, \textit{Quantum Statistical Properties of Radiation} (New York, 1990), Chap. 1, Eq. (1.16.23).

\end{thebibliography}
\end{document}